# Is the brain relativistic?
D. Le Bihan[1,2,3]
*denis.lebihan@gmail.com*
[1]**NeuroSpin, CEA-Saclay Center, Gif-sur-Yvette, France**
[2]**National Institute for Physiological Sciences (NIPS), Okazaki, Japan**
[3]**Human Brain Research Center, Kyoto University, Kyoto, Japan**
(*Dated: August 9th, 2019*
*Revised Sep 1st 2019*))

**Abstract**

Considering the very large body of knowledge which neuroimaging has put at our fingertips over the last three decades we looked at the brain with a fresh view which could unveil those "old" things in new ways, in a framework which could help us making predictions tailored made for a scrutiny with the outstanding imaging instruments to come, such as ultra-high field MRI. By doing so, switching back and forth between physics and neurobiology, we came across the view that time and space in the brain, as in the Universe, were, indeed, tightly mingled, and could fade away to be unified through a brain "spacetime". Considering that there is a speed limit for action potentials flowing along myelinated axons further thinking led us to envision that this 4-dimensional brain spacetime would obey a kind of relativistic pseudo-diffusion principle and present a functional curvature governed by brain activity, in a similar way gravitational masses give our 4-dimensional Universe spacetime its curvature. We then looked at how this whole-brain framework may shed light on brain dysfunction phenotypes (clinical expression of diseases) observed in some neuropsychiatric and consciousness disorders.

## 1. Introduction

The recent publication of the first image of a Universe black hole, an immaterial spacetime structure, as seen by the Event Horizon Telescope (EHT) [1] has triggered a lot of attention worldwide. This applause is largely well deserved given the technical, instrumental and data processing tour de force it represents, but also because it is a major landmark which epitomizes the features of the General Relativity theory (dynamics of curved spacetime) outlined by Einstein a century ago [2]. Though there are still many question marks and inconsistencies to resolve to understand how matter, antimatter, radiation, dark matter and dark energy contribute to our Universe, physicists have at their disposal an array of models and gauge theories, (eg the Standard Model, Quantum Electrodynamics, etc.) and instruments (eg EHT, or the Large Hadron Collider (LHC) at CERN) to evaluate them.

In comparison, neuroscientists are still lacking a global theory of the working brain as a whole which could give accounts of cognition, behavior, up to our mental life or else consciousness, although we know a lot at molecular and cellular level about the working of individual neurons that make our brain (about $10^{11}$ packed in gray matter, roughly the same as the number of visible stars in our galaxy) and their connections through white matter. Could we dispose one day of a Gauge Theory of the brain? This goal could be seen as one of the main scientific challenges for the 21$^{st}$ century with obvious implications for our society at large, allowing us to better understand and cure neurological or psychiatric disorders, or improve our cognitive performances, especially in aging populations, with obvious economical counterparts. In some way neuroscientists now also have "instruments": Neuroimaging has become an inescapable pathway, because it allows getting maps of brain structure and function *in situ*, non-invasively, in patients or normal volunteers of any age. Magnetic Resonance Imaging (MRI), especially, has become the reference approach to investigate the human brain functional anatomy *in vivo* because it provides high resolution images without ionizing radiations or the need for tracer injections. The next decade will see ultra-high field MRI systems (with magnetic field operating above 10 teslas, [3]) becoming operational, providing a yet unseen view of our brain at a boosted spatial and temporal resolution.

Such revolutionary instruments will allow us, hopefully, to discover new, unpredicted features, as has been the case, for instance, after the invention of the microscope. But those instruments will first be used to confirm or invalidate current assumptions or theories, as a crucial requirement for a theory is falsifiability, the possibility of experimental disproof. Considering the very large body of knowledge which neuroimaging has put at our fingertips over the last 3 decades we thought that we could perhaps look at the brain with a fresh view which could unveil those "old" things in new ways, in a framework which could help us making predictions tailored made for a scrutiny with the outstanding instruments to come. By doing so, switching back and forth between physics and neurobiology, we came across the view that time and space in the brain, as in the Universe, were, indeed, tightly mingled, and could fade away to be unified through a brain "spacetime". Further thinking led us to realize that this 4-dimensional brain spacetime would obey a kind of relativistic principle (obviously different from that prevailing for the physical Universe) and present a functional curvature generated by brain activity, in a similar way gravitational masses give our 4-dimensional Universe spacetime its curvature. We then looked at how this whole-brain framework may shed light on brain dysfunction phenotypes (clinical expression of diseases) observed in some neuropsychiatric and consciousness disorders.

## 2. A brief account of our current knowledge of the brain
### 2.1. The tools: Neuroimaging

Much important progress has been made using neuroimaging methods which allow to investigate the whole brain of human subjects non-invasively. While plain MRI reveals exquisite details about brain's anatomy, functional MRI (fMRI) has emerged as an important approach to study the brain and the mind, as it reveals activation status of neurons assemblies in brain gray matter in various situations [4]. On the other hand Diffusion MRI [5] which refers to imaging of the random translational motion of molecules (also called Brownian motion), a physical process which was first characterized by Einstein [6], provides information on brain tissue microstructure at cellular scale (a kind of virtual microscopy), both in brain gray and white matter [7]. In white matter which consists in connective fibers (axons) emanating from neurons water diffusion was found anisotropic, direction dependent. This feature can be exploited by Diffusion Tensor Imaging [8-9] to produce stunning maps of the orientation in space of the white matter tracks and brain connections, as well as to provide information on white track microstructure and integrity. The potential of diffusion MRI and DTI to probe human brain connectivity has attracted worldwide interest. The European FP7 CONNECT project and Human Brain Project as well as the Human Connectome Project have clearly benefited of those powerful approaches, yielding insight into how brain connections underlie function. DTI which is commonly used to investigate white matter disorders has revealed faulty brain connections linked to dyslexia, schizophrenia, autism, bipolar or anxiety disorders, suggesting that disconnection may alter brain processing in patients with those disorders.

Functional connectivity between brain areas is usually estimated by calculating the relationship between regional time series using correlations, mutual information or coherence. Functional connectivity maps can be obtained either in the context of task-related brain activity through fMRI or in "resting" conditions through resting-state fMRI (rs-fMRI) [10]. The first drafts of the human "connectome" were made by combining results of DTI, fMRI and rs-fMRI in large cohorts of normal subjects, although rs-fMRI is subject to important potential confounds which have to be taken into account during processing. Furthermore, the relatively poor spatiotemporal resolution (typically millimeters and seconds) and the indirect nature of the imaging markers to reflect neural activity remain important limitations which can be partly overcome by combining the methods (eg spatial resolution of MRI and temporal resolution of magneto-encephalography, MEG), or completing knowledge by more invasive methods performed in animal models or neurosurgical patients (eg electrophysiological recordings).

### 2.2. The nodes and edges of the brain

The brain is a spatially very inhomogeneous organ. Following the postulate made by Broca [11] that the brain was organized along functional regions our contemporary vision of the brain obviously results from Cajal's discovery that neurons do not constitute a single reticulum but are individual units with interconnections [12]. Neurons are disposed in the brain (mainly in the basal ganglia and the cortex) following a multiscale architecture at all levels (molecular, micro-, meso- and macroscopic). The segregation of cells in a set of functional areas along the cortical surface separated with abrupt boundaries has been known since Brodmann gave an account of his observation under a microscope of the single half-brain at the onset of the 20th century [13], and found out the existence of about 60 distinct areas which he labeled by numbers (for instance, area 17 is the primary visual cortex). Later those areas were found to have also a chemoarchitecture signature [14]. A recent MRI study has extended this set of brain regions to almost 200 [15]. While this structural and functional architecture has undeniably a genetic basis [16] a key question is to understand how the specific three-dimensional organization of the brain cells, neurons and glial cells, in clusters or networks within the layers amd columns of the brain cortex, and their short- and long-term *dynamic interactions* via their short and long range *connections*, are responsible for the emergence of a set of elementary operations which, combined together and under the effect of exposure to environment result in higher order function, language, calculus, or even consciousness. Infants can today learn to manipulate cell phones with some success in a few hours, though, there are certainly no "cell phone" brain areas, nor to say "cell phone" genes.

In some way, we are in a situation conceptually equivalent to that which led to the discovery of the genetic code. The existence of our 46 chromosomes and their link to heredity was revealed in the mid-1880s. At a more microscopic level it was later discovered that the vector of heredity in chromosomes was the DNA molecule, an assembly of nucleotides. For the brain we know that it contents a finite number of functional areas, which are made of neurons which carry brain function. But it was not until Crick and Watson had the intuition that information was hidden at an intermediate spatial scale, between DNA's nucleotides and chromosomes scales, in the three-dimensional organization (double helix) of the DNA molecule that the existence of the genetic code emerged [17]. Could there be a "neural code" carried by the three-dimensional organization of brain cells within each cluster? The very specific organization of the visual cortex in layers and columns giving this cortex its unique function was revealed by Hubel and Wiesel [18]. It remains to be seen whether this layer/column organization in ubiquitous in the brain, structurally and functionally. Obviously, because such precise organization must have a genetic origin, only a fixed set ("alphabet") of specific low level functions are expressed. Over learning and environmental pressure modulation of those areas might occur in terms of enhancement or deactivation (as demonstrated with the temporal windows present during the early stages of development). However new basic functions cannot appear. For instance, humans cannot read QR codes (designed for computer algorithms to access web sites from a visual input), even after extensive learning, because we do not have the proper circuits, although humans designated the algorithms to produce those QR "images".

Higher order functions then result from the dynamic combinations of those "alphabet" low level functions, which implies connections between those neuron clusters or "nodes". Interestingly, clusters of neurons with similar functions tend to have similar connections patterns and similar gene expression profiles, especially in highly interconnected hubs [19]. Curiously, the crucial importance of those connections has only relatively recently emerged, in part due to the arrival of DTI [9], which has given brain white matter an interesting revival, through the noble term of "edges" to borrow the network jargon. The brain is now seen as a global network of nodes of different sizes and configurations connected through "hubs" [20]. However, what may have not been perhaps fully realized yet (probably because we do not dispose of instruments to directly investigate it) is how neural activity flows along those connectivity lines. While *spatial* encoding appears as the landmark feature to explain "gray matter" and how neural cluster specific activities arise from the unique architecture of their cell constituents, *temporal* encoding seems the foremost feature embedded in "white matter", as transmission of information between nodes necessarily implies time delays and issues with coherence and synchronization among nodes. By space and time we do not refer here to our perception of those physical world features, but to how the brain internally appropriates time and space to carry its "function".

### 2.3. Current global brain function models

At the very fundamental level action potentials represent the most basic form of communication utilized by neurons, as set by the Hodgkin and Huxley model [22]. Information that are received and subsequently transmitted from a neuron from/to its neighbors are varied through the temporal domain (eg frequency and phase). As we traverse from the scale of neurons to microcircuits then brain networks, temporal patterns remain the predominant mode for how information is transmitted and propagated [23]. However, the situation becomes less clear when considering the brain as a large scale system integrating collective neural activity, as there is yet no broadly accepted mathematical or physical theory making consensus. This scale, however, is critical in supporting high level brain activity, such as cognition or consciousness, and their diseased counterparts, as with mental disorders. Ideally one would like to dispose of a model or framework which allows for prediction, testing and possibly rejection.

Using graph theory in combination with DTI and rs-fMRI data a brain network of nodes (parcellation of gray matter in a finite number of functional clusters, typically a few hundreds) and connections ("edges") can be built. The brain can then be considered as a "small world network" structured around a large number of spatially distributed network communities with highly segregated computation capacities, integrated through network hubs, making a network topology [24]. Connectivity level is estimated from the strength of connections between the nodes, but the view is in general only descriptive, lacking the dynamics of brain functional activity. In all cases one should consider the spatiotemporal architecture of the neural systems and their physical constraints, as they clearly affect neural function.

With whole-brain computational modeling [25], under the dynamical systems theory, sets of system's differential equations can be linked to brain state variables (derived from biophysics of neurons, eg, flowing ions and membrane potentials) in a phase space, benefiting from noise and stochastic calculus. Grossly assuming that a large spatial scales neuronal activity becomes linear and not correlated (normally distributed) the Fokker-Planck equation may be used to build a Neuronal Ensemble reduction model. In another approach, taking into account strong coherences between neurons, a further reduction can be made through Neural Mass Models. Large brain dynamics can then de declined into a discrete network of neural masses or nodes (mesoscopic circuits) interacting with distal regions through the connectome. Neural field modes, on the other hand, consider the cortex as a continuous sheet with dense short-range connections.

Regrettably, despite promising results for clinical applications and thought-provoking reviews [26-29], those innovative models have not well permeated the field of neuroscience, probably for technical and cultural reasons. It surely takes a malleable mind to navigate through concepts inherited from disciplines as far apart as physics, statistical mechanics, mathematics, information theory, neurobiology, psychiatry or even philosophy to grasp the essence of those models. Nevertheless, interdisciplinarity is the key to understand some day the secrets of our mind. Some mathematical models based on network information theory may also appear to some neuroscientists as mainly phenomenological, lacking physical or even biological content. Indeed, a major difficulty when building a large-scale neuroscience model is to rely on abstract physical and mathematical tools to allow predictions without sacrificing too many neuroanatomical, physiological and biophysical details (though they may become fuzzy and less relevant with upscaling) to remain realistic and close to experimental results. However, both microscopic views (even at molecular level) and coarse-grained approximations (more accessible to neuroimaging) are not at all exclusive. In fact they are necessary and complementary, given the multiscale structure-function organization of the brain, so that both biophysical realism and functional phenomenology have room to coexist, and we must accept that what might be seen as truth for one may well be perceived as speculation by another. The border will become even more indistinct with the next generation of neuroimaging instruments which will boost spatial and temporal resolution and bridge scales.

## 3. Toward a relativistic brain framework
### 3.1. Relativity of time and simultaneity in the brain

There is an immense body of literature (scientific and philosophical) on the concept of time, and it is out of the scope of this short essay to review or humbly comment the many viewpoints which have emerged in the past (some of which are centuries old) from such giants as Aristotle, Saint Augustin, Newton, Leibniz (who even dropped the letter 't' from his name as he did not believe in the absolute time, t), Newton, Kant or else Bergson. At this point let us take a perspective on how our brain is connected to the physical world through our body.

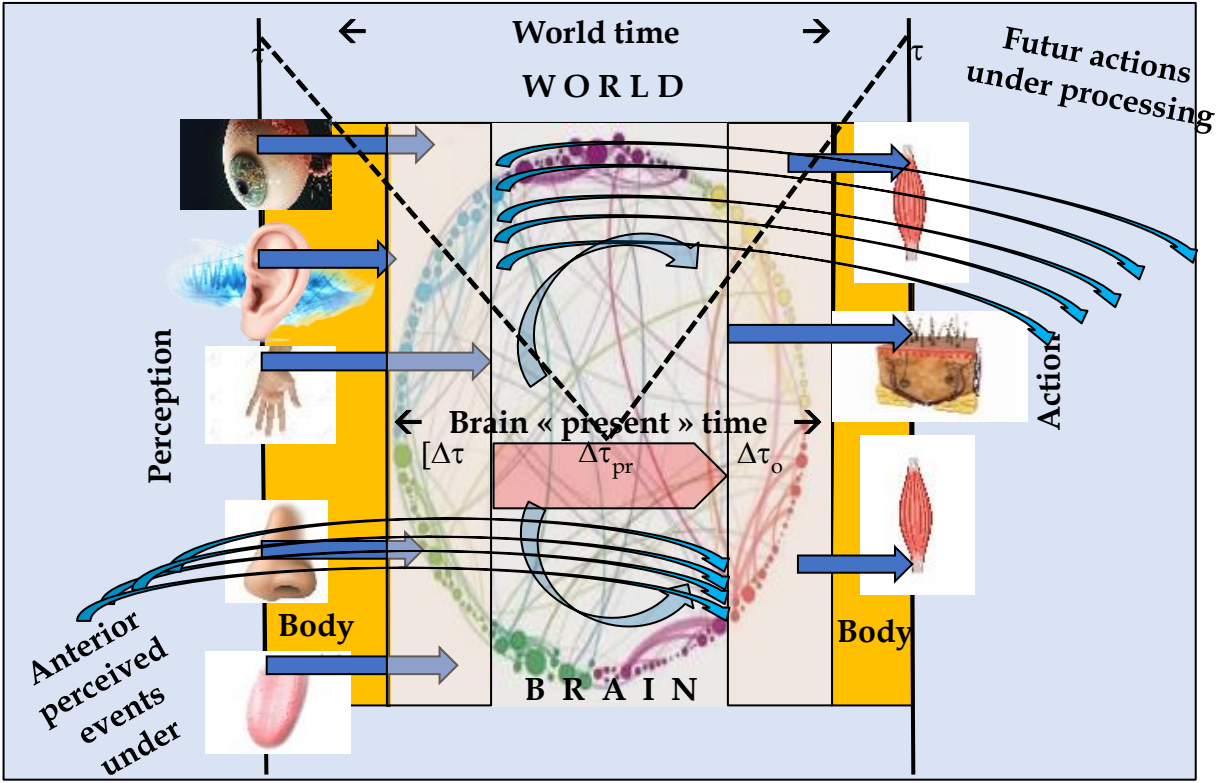

**Figure 1: Time is ill-defined in the brain**
*The brain interacts with the world indirectly through the body which results in time delays between perception and action. Beside brain processing adds up its own time, especially considering that past events (memory) and storage for future actions are combined to produce the best responses to environmental stimuli.*

The mind of even the finest intellectuals, philosophers or artists comes to nothing if it cannot communicate to others, which implies using the body in one way or another. Suffice to say that, for simplicity, that the brain "perceives" the world through the body which is upholstered with a very great number of sensors, externally (giving our senses, vision, audition, touch, taste, odor), as well as internally (Fig.1). Reciprocally, the brain "acts" on the physical world through the body (also externally and internally), mainly via muscles which allow "action" or chemical release (through action on glands). Sensors and actors are linked to the brain through peripheral nerves which have a limited transmission speed, so that transmission cannot be instantaneous. Furthermore, between perception and action some processing occurs in the brain which adds to the minimum time required for an action event to reply to a perception event. This delay can be very short (taking as an example the movements of the fingers of a piano player interpreting a piece), but also very long (taking as another example the movements of the fingers of a shogi or chess player responding to a move from the opponent). Indeed, brain processing spreads over several time scales, as information perceived is mixed up with prior "knowledge" acquired during our brain history (which obviously starts well before our birth) and stored as "memory".

Similarly, resulting actions may be spread in time (the shogi player deploying a strategy is again a good example). Those facts should take us to consider that the time which elapses internally within our brain might differ from the physical world time, and, indeed, to question whether the "present" concept exists at all in our brain. It is at least ill defined (we are not referring here at how the brain *perceives* time, on which there is a very abundant literature, see for instance [30]) as past and future are intrinsically intermingled across various scales. Because of the unavoidable delays between perception and action the brain must anticipate through predictive models in order to respond adequately to its environment (either by modifying it or by modifying oneself). In other word, the brain makes a *future* image of the world to *act* on the *past* images it *perceived*. While the physical world absolute time concept was abolished as a consequence of the relativity theory we might also have to consider that there is no absolute time in our brain and that the brain time is entangled with the brain spatial organization, from microscopic to macroscopic scales, which builds up upon its entire lifespan.

### 3.2. The existence of a speed limit in the brain connectome

At the microscopic level, in the mammalian nervous system, conduction delays along one axon vary widely, from less than 100 microseconds to more than 100ms. Those delays depend, obviously, on the axon length, but, counterintuitively, more on the axon diameter. Axons larger than 0.3 micrometers are generally myelinated, that is surrounded by a spiral of lipid layers produced by glial cells (oligodendrocytes). While the myelin sheet can be seen as an electric insulator around the axon, it gives it, indeed, its speed: There is a direct relationship between the conduction velocity, the axon diameter and the thickness of the myelin (the ratio of the myelinated fiber diameter to the axon diameter is called the g-ratio, closed to 0.6)[31]. Myelinated axons of 20 micrometers in diameter conduct at a speed of >100m/s while short unmyelinated axons conduct at 0.3m/s. The reason for this speed/diameter relationship is that the myelin sheet is interrupted at regular intervals (which depends on the sheet thickness). Those interruptions, called Ranvier's nodes, make the ion channels present on the axonal membrane exposed to the extra-cellular space only at those locations, so that electric currents and potentials literally jump between nodes, instead of propagating continuously along the axonal membrane, making this "saltatory conduction" (in fact a regeneration of action potential from node to node) a very efficient mechanism to speed up conduction velocity (and reduce "latencies"). Another important feature is that fibers have a "refractory period": Once an action potential has passed at a given location no other one can follow for a short time interval. This feature also prevents the action potentials to travel backwards, the axon being a one-way path, but also limit the amount of action potentials (and, hence, information "throughput") which can be carried by the axon by unit of time.

Space is a premium in our brain so all axons cannot benefit from large myelination, or the size of our head would have to become prohibitive and difficult to carry (not to mention the associated logistic issues, especially energy distribution through blood flow). Hence, the most myelinated fibers are those which are the longest (like those emanating from the motor cortex in the brain toward the bottom of the spinal cord where they connect to the motoneurons, giving us control of our limb muscles) to save on conduction times. Also, myelinated fibers are grouped within bundles of fibers connecting hubs of gray matter. The size of those bundles determines how much information can be carried ("bandwidth"). As in an electrical network the latency (which, for the axon, is inversely linked to the conduction velocity) is the ultimate parameter which determines the amount of data which can be transferred over a time period. Latencies vary greatly depending on axon locations, from a few to over 100 milliseconds in primates. Interestingly, conduction velocities may also fluctuate according to the activation status along the fiber or over time (supernormal or subnormal conduction, [32]). Of course, latencies can be dramatically altered in relation to neurological diseases (such as Multiple Sclerosis), but also in some psychiatric disorders, as recently evidenced by DTI (see below).

In short, considering that neural activity flows sequentially along multiple axons it is not overstated to admit that the brain is a slow organ. In any case, the brain conduction speed limit (let us call it $c^*$) is, thus, an essential (but often forgotten) fact to consider when modeling brain function, not to mention synchronization between highly connected brain areas, pointing out again to the ill-defined nature of the "present" concept within our brain.

### 3.3. How the brain optimizes its speed limit, $c^*$

Earlier studies have postulated that energy-savvy wiring minimization is a fundamental driver of the connectome [33-34]. According to this model one immediately sees that there is an evolutionary advantage (in terms of calculation power) for the brain to dispose of the highest possible speed limit, which is gained by myelination of the axons constituting brain white matter. Considering the humain brain this increase is obtained through the myelinization process which is part of the brain maturation taking place early in life until the end of puberty or even later for some white matter tacks. At birth the brain with its capital of under $10^{11}$ neurons weights around 400g, each neuron being connected to about 500 other neurons in average. Yet after adolescence the brain weight will reach about 1.3kg although the number of neurons remains the same (or slightly less). The explanation lies in the increase of connections, each neuron being connected to about 10000 other neurons in average, and myelination of axons, which underlines the importance (until now neglected) of white matter and the connectome. In the visual system, for instance, the P1 peak latency observed with evoked potentials in the visual areas following visual stimulation decreases from almost 300ms at birth to 100ms after 10 weeks post terms [35] (Fig.2). In parallel the Fractional Anisotropy (rough index of axon myelination obtained with Diffusion Tensor Imaging) in the optic radiations fibers increases in about the same ratio [36] (Fig.2).

However, efficient communication may also be obtained through another mechanism somewhat complementing brain wiring minimization [37-38]. Synaptic pruning is the process which eliminates synapses starting near the time of birth and continuing into the mid-20s. Pruning is determined by exposure to environment and learning: At birth, many potential connections are possible and will then be tested and consolidated depending on our experiences, often at random (Fig.3).

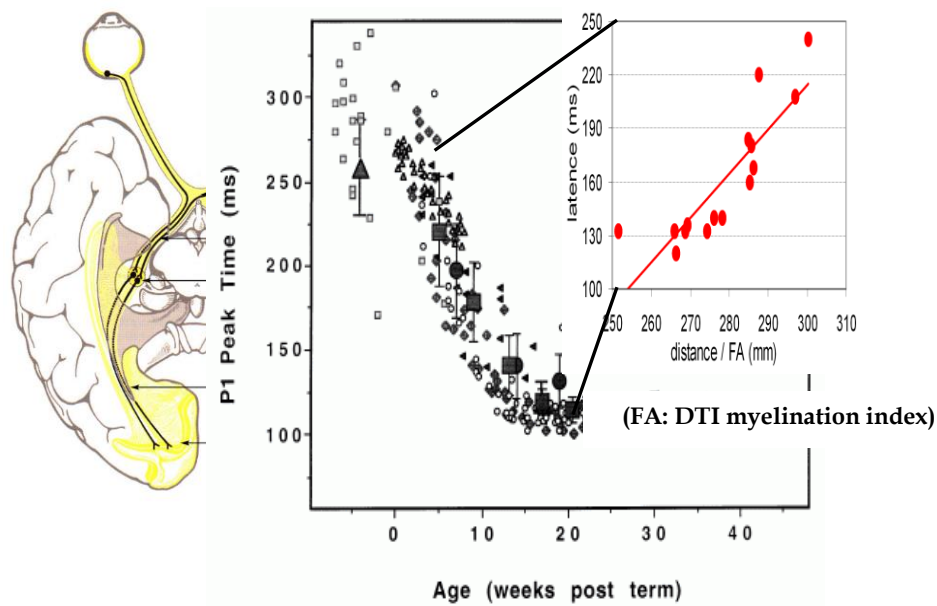

(FA: DTI myelination index)

**Figure 2: Maturation of visual system**
*The time required for visual inputs to reach brain early visual areas (as estimated from Evoked Potential Latencies) decreases when brain matures at an early age. The increase in transmission speed results mainly from the myelinization of the related white matter tracks (optic radiation), as revealed through the increase of water diffusion fractional anisotropy (FA) measured with DTI (adapted from [35] and [36]).*

Over time connections which are not deemed "necessary" are eliminated, leaving only connections that are made with the functionally appropriate processing centers. This is the reason why, for instance, Japanese adults have difficulties separating the sounds "r" and "l" (as in "election" and "erection") which are not present in the Japanese language, while Japanese babies have no problem under about 1 year of age [39]. By removing unnecessary pathways this mechanism will speed up the "networking" capacity within the brain and increase its efficiency (Fig.3). This mechanism also occurs dynamically through learning, resulting in some kind of "binding" within and between neural circuits [40], and gives roots to memory [41]. Indeed, experience-dependent changes in adult brain white matter (intraparietal sulcus) have been demonstrated in adults following training of a complex visuo-motor skill [42]. On the other hand, if pruning and synaptic consolidation increase performance for some tasks, they also somewhat increase our brain "stiffness" and to some extent reduce our potential to learn by depriving us to see the world with "child eyes".

In the end, brain maturation through myelination and pruning will make structural hubs to emerge naturally, surrounded by edges tuned to optimize network capacities. Hence, it appears that both wiring optimization and communication efficiency contribute highly to the development of brain network organization through spatial embedding. Interestingly, Nadkarni et al. [43] have shown that cortical to cerebral volume ratios were similar in all primates (marmoset, macaque, humans) including mouse lemurs, and lower in mice. However, white matter to cerebral volume ratios increased from rodents to primates (mouse lemurs and marmosets) to macaque, reaching their highest values in humans. In remains now to see how time can also be embedded in this 3D spatial network to constitute a brain 4D spacetime.

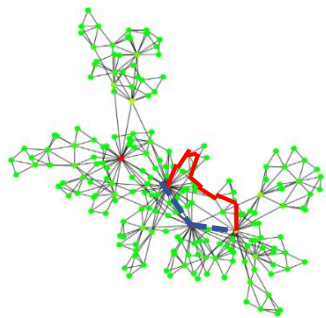

**Figure 3: Synaptic pruning**
*Over the first years of life a great number of synapses are formed between neurons, but many will not survive. Selection is made through exposure to environment. As a result, connections become simplified and pathways more efficient (shortest paths).*

### 4. The brain spacetime

Following the arguments developed above one should not find it objectionable, we hope, that the brain may be viewed in some way as a physical "object" embedded in a 4D enclosure. While the brain which is part of the Universe must obey Universe physical laws one may also consider that the brain represents a kind of Universe itself where physical laws could be revisited, directly or imported through analogical derivations to provide a framework useful to better represent and perhaps understand how the brain works as a whole system. After all, the perception we have of the external world, the Universe, comes from our internal world, that is our mind in our brain, and it should not come as a surprise that our understanding of the Universe and our brain are irremediably connected.

Interestingly, while nearly all fields of physics have contributed to our understanding of the brain (directly or through the neuroimaging instruments to explore it), from thermodynamics to electromagnetism, relativity had not been considered (except for the subjective perception of time [44]), obviously because the light speed limit has not been seen as an issue in the brain. This is certainly true, until we appreciate that the brain carries its own speed limit. Given the above considerations about time issues in the brain we examined a new paradigm where brain function would no longer be described in terms of absolute space (brain anatomy and connections) and absolute time (times delays along connections), which have also be considered by some authors [45], but rather through a *combined* brain spacetime, to mirror the physical spacetime which emerged from the special relativity theory. This new framework emerged from the ascertainment that there is likewise a speed limit in our brain, obviously much slower than the speed of light, but nonetheless a limit which is not without consequences as we shall see. This view led us to consider brain geodesics and an associated metric determining how neural "activity" (action potentials) flows within the brain, linking network (functional) topology with network (structural) geometry in a 4D space. Space and time are completely entangled in the brain, more than in the physical Universe, as temporal synchronization plays a crucial role in the functioning of the brain. The brain is sensitive to changes in is environment, resulting in fluctuating internal patterns, and should be described more in differential terms than continuity.

Obviously, fundamental differences exist between the brain and the Universe, so analogies cannot be pursued too far. For instance, for cosmologists the Universe is considered to be infinite, isotropic and homogeneous. To the opposite the brain is a spatially very inhomogeneous organ with a complex organization. Nonetheless, the Universe also presents some organization (galaxies, clusters and superclusters forming an irregular lattice), and all depend on the scale we consider. The brain which is obviously not infinite (we will see the consequence of this trivial fact at the end of this essay), sees its "speed limit" evolve over time, as mentioned above, during normal brain maturation or the course of a disease, while the Universe speed limit (in vacuum) is considered constant is space and time (however, some authors have suggested that the speed of light might have been different during the infancy of our Universe, just after the big bang [46], or varied in time in the context of the Hubble Universe expansion model to explain the red shift of distant galaxies [47]). Also the configuration of our brain spacetime is expected to changing dynamically extremely fast, while the Universe evolution (mainly its current expansion) is extremely slow, at least to the Human scale. Finally, our Universe is considered 'healthy' while our brain may, unfortunately, suffer from many disorders affecting its physical properties.

### 4.1 Brain spacetime metric

The existence of a speed limit implies that nodes and signals traveling between them have their own, independent time frames. It should be noted that the existence of a speed limit does not imply, however, that signals all travel at this speed, far from there. Whether anything physical actually travels at this speed limit is really irrelevant for the logic of the model. But simultaneity becomes ill defined. Events which are simultaneous in a given frame do not necessarily occur at the same time in another frame. What is important is not the time per se, but changes, the fact that the brain works as a network of events (we may also say processes) which are interrelated, and not in a passive way around immutable constituents.

It might be important to distinguish between the sets of events that a brain node *sees* at one instant and the set of events that the node considers *having occurred* at that instant. What a given node actually sees at one instant is a brain picture. It is a composite of events that occurred progressively earlier as they occurred farther away. This picture represents all events which could have influenced the node activity *now*. On the other hand, we may also consider the *brain map* that we see as an external observer, a 3D map of events occurring in all nodes at $t=t_0$, a kind of frozen instant global map. This brain map is what matters to us because this is what is accessible to our instruments. But what one node actually "sees" may be rather different.

We introduce here a new framework for combining space and time in the brain borrowing a concept inherited from Minkowski's spacetime [48] to depict Einstein's special relativity features [49]. This frame allows the standard *static* picture of the brain "nodes" and their "connections" to be replaced by a *dynamic* picture. To do so we focus on brain 'events', the 'atoms' of our brain history (which starts as soon as neurons get connected, hence well before we are born). With this new approach brain events are represented as points of a 4-dimensional brain spacetime where space and time are symmetric, with (c*t) appearing as a spatial dimension. In other word time and space get mixed to a degree which depends on c* and a relativistic factor γ which is always smaller than 1. In this spacetime one may consider an "active" view where events P(r,t) move to a new spacetime position Q(r',t') in the same frame. With a "passive" view P(r',t') is merely a new label of the event P(r,t), as if only coordinates change. To the brain spacetime we associate a 4-dimensional metric in quadratic differential form to assign "distances":

$$ds^2 := c^{*2}dt^2 - dx^2 - dy^2 - dz^2 \qquad [1]$$

where $dr^2=dx^2+dy^2+dz^2$ corresponds brain nodes "spatial" coordinates (in nodes space), or more generally, to follow Einstein's tensorial notation:

$$ds^2 = g_{\mu\nu} dx^\mu dx^\nu \qquad [2]$$

where $\mu,\nu = 0, 1, 2, 3$ such that $x^\mu = \{c^*t, x, y, z\}$ and $g_{\mu\nu}$ is a symmetric metric tensor with $g_{\mu\nu} = \text{diag}(1, -1, -1, -1) = g_{\nu\mu}$.

Brain activity will flow in the brain spacetime following geodesics, i.e., "straight" lines in this pseudo-Riemannian space. The "path length" concept frequently referred to with network models as the shortest path between two nodes will now have to be considered in this 4D spacetime geometry, the time dimension becoming a part of the path. Hence, shortest paths may no longer be associated with shortest physical distances. So there is no single time, there is a different duration for each path, and time passes differently according to speed and location.

This brain spacetime provides a fresh physical framework to explain our mind, of heuristic value, not just a mere artifice. First, it gives us another vision, namely looking at brain spacetime from a traveling action potential perspective (with nodes appearing as sequential landscapes along paths) and no longer from a static node point of view (with arriving and departing action potentials). As a comparison, we are now looking at flight maps from within an airplane and no longer from an airport. Second, as we will see later, this spacetime can be acted upon by the brain itself which distorts it, giving it curvature, as gravity does to physical spacetime under general relativity.

### 4.2 Brain Minskowski diagram

As special relativity was 'geometrized' with Minkowski diagrams we may also visualize spacetime relations in the brain using similar diagrams. To map such 4D spacetime on a piece of paper of 2-dimensional (Euclidian perspective rendering) figures one has to distort metric relations, such as the unavoidable absence of 1 or 2 dimensions, and mentally compensate for this distortion, for instance reading circles as spheres. Figure4 shows an example of such rendering around a diagram. The vertical axis corresponds to time. Brainlines, representing brain activity "travelling" between successive events subtend angles less than a maximum value with the vertical, since the slope relative to the axis is fixed by the maximum speed allowed by the network.

The most fundamental structure in this spacetime is, indeed, the set of "cones" at each point (event) P. They represent the bundles of brainlines of all activity passing through P at maximum potential speed, or, equivalently, the loci of events that can send activity to or receive activity from P at this speed, or yet, the loci of events at "zero interval" (in 4D-space) from P: $\Delta s^2 = 0 = c^{*2}\Delta t^2 - \Delta x^2 - \Delta y^2 - \Delta z^2$. Under the suppression of the z-dimension this 4D-space reduced to a cone figure representing a front in the spatial xy-plane converging onto the origin and then diverging away from it. In full dimension it is not a circular, but a spherical front that converges onto and diverges from the event. Those cones at each event can be seen as grains in the brain spacetime, with the brainlines having to be within the cone of each of their points (Fig.4). By formally letting the maximum speed to be infinite and using $(c*t)^2$ for the "time" axis the cones become flat: The brainspace reduces to a 3D-Euclidian space where time and simultaneity are identical for all nodes (standard static picture of the brain).

Let us now consider 2 events (not necessarily neighboring) P and Q. The "distance" between those events is $\Delta s^2 = c^{*2}\Delta t^2 - \Delta x^2 - \Delta y^2 - \Delta z^2 = \Delta t^2(c^{*2} - \Delta r^2/\Delta t^2)$. If $\Delta s^2 = 0$ we have seen that P and Q are connectible by signals travelling at the maximum network speed limit. When $\Delta s^2 > 0$ one has $\Delta r^2/\Delta t^2 < c^{*2}$ and signals can be exchanged between P and Q. In the particular case P and Q occur at the same location ($\Delta r^2 = 0$) one has $\Delta s^2 = c^{*2}\Delta t^2$. The interval $\Delta s^2$ is simply $c^{*2}$ times the time elapsed between those 2 (connected) events occurring at this location (for instance, through a feed-back loop). When $\Delta s^2 < 0$ one has $\Delta r^2/\Delta t^2 > c^{*2}$ which would imply activity to travel *faster* than the network speed limit. Taking $\Delta t = 0$ one gets $\Delta s^2 = -\Delta r^2$ which corresponds to the shortest possible spatial separation between to P and Q taken as simultaneous events.

Brain activity will thus follow geodesics within each successive nodes cone, making a line of possible events (we call "brainlines"). Other lines are "forbidden", as events occurring in other nodes at the same time ("present" defined for this node event) are out of reach. An event which arrived from the "past" at a given node from another node can go back to this node in the "future", making a backward projection (loop). Several brain lines can "meet" at a given event to act together in the node within synchronization time windows.

The cones at each event P set a very important causal partition of all other events relative to P (Fig.4). All events on and within the *future* cone (top half) can be influenced by P; for they can receive signals from P. All events occurring in this region happen *after* P and constitute the

*absolute (causal) future* of P. Conversely, events on or within P's past cone (bottom half) *precede* P and constitute the *absolute (causal) past* of P. No event in the region outside the cone (Δs²<0) can influence P or be influenced by it, but can be simultaneous with P, hence this region corresponds to the *causal present* of P.

Another interesting consequence of this view is that clusters of brain nodes may not be considered as "rigid bodies" within the brain 4D spacetime. The geometrical shape of those clusters in this spacetime will evolve continuously in time depending on the speed experienced among the brainlines connecting them. Similarly, the features (length, frequency, etc.) of trains of action potentials "traveling" along those brainlines will vary, changing their time profile.

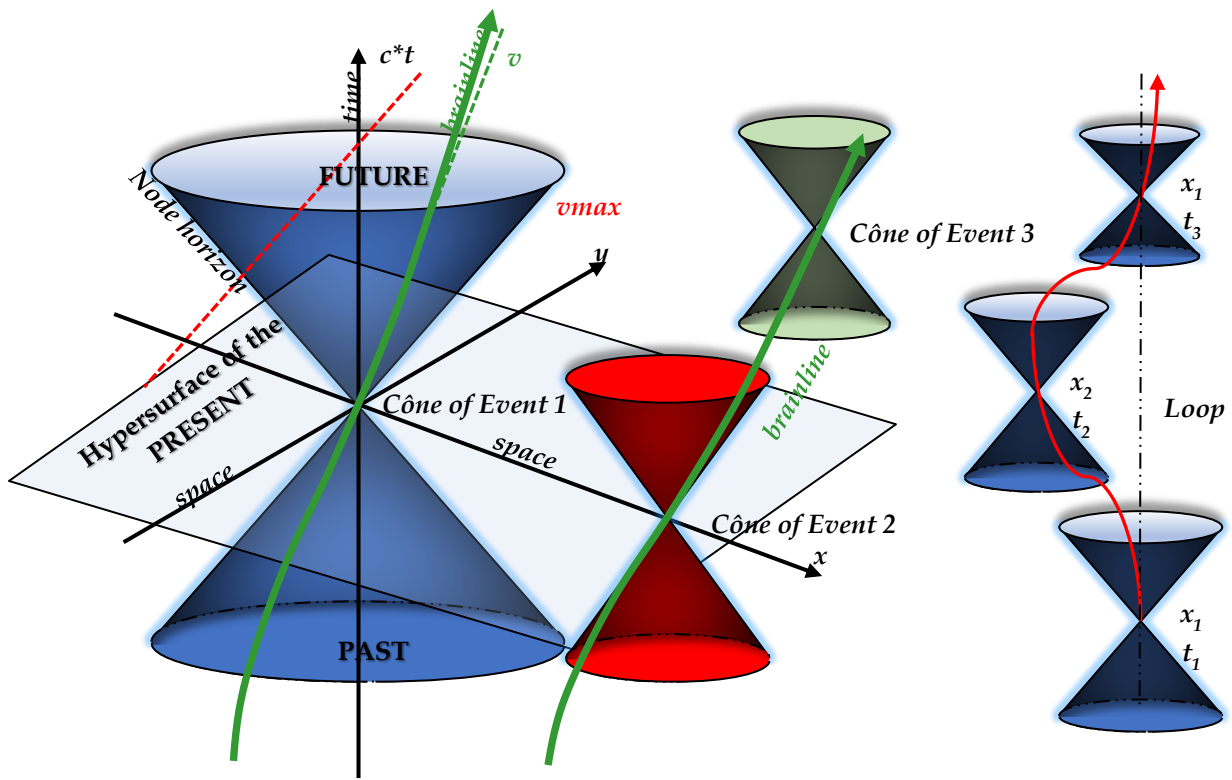

**Figure 4: Brain spacetime Minkowski diagram**

*The Minkowski diagram was introduced in 1908 to provide a straightforward view of the relativity theory, reducing spacetime to 3 dimensions instead of 4. At a given "event" in spacetime a cone is determined by the speed limit in the spacetime. Events are connected in spacetime through timelines (here brainlines). For a given event only brainlines which remain within the cone of the event are causally related (in the past or the future). Events occurring simultaneously for this event cannot be connected (hypersurface of the present). A brainline passing through the same location overtime represents a loop.*

### 4.3 Application to "disconnection" disorders

Schizophrenia has often been considered as a "disconnection" disorder [50-51] and many studies, mainly using diffusion MRI and DTI have revealed specific network alterations, especially reduced structural connectivity [52] between frontal, temporal areas and multimodal cortices. The origin of the disease (in terms of the reason for those abnormalities to exist, genetic factors, developmental defects especially over pruning in childhood [53] remains a subject of debate, but the phenotype of the disease (clinical symptoms) might be explained using the Minkowski diagram (Fig.5). As the wiring becomes less efficient the speed limit may become unstable and often reduced along those tracts, while it remains normal along other tracks. Thus, it may happen in some instances that activity flows at a "supraliminal speed" with regards to the faulty connections. As a result the order of events among "successive" nodes in the 4D brain spacetime may be altered or even reversed.

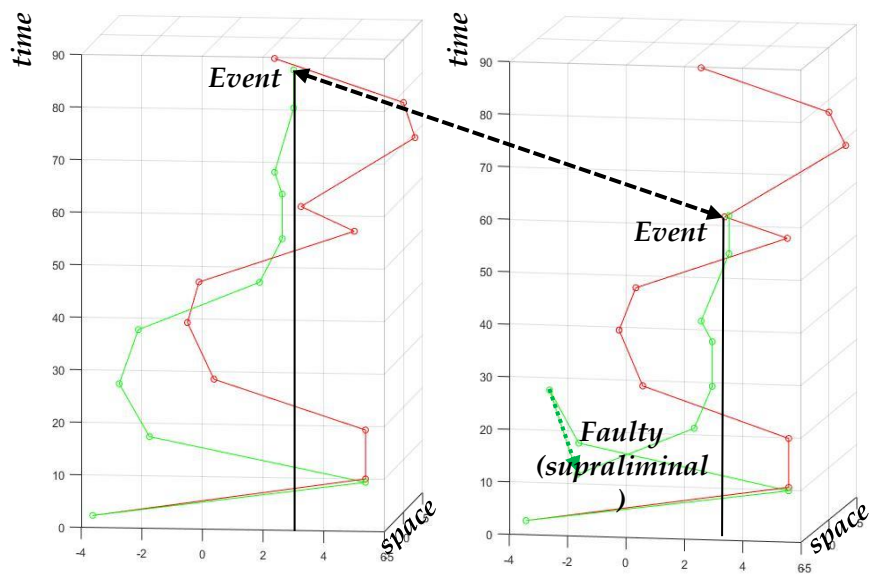

**Figure 5a: Reversal of event time order in schizophrenia**
*This left diagram represents 2 brainlines in brain spacetime originating from the same event but following different geodesics ending up in a close spatial location after 90ms. In the right diagram a faulty connection has been introduced along the pathway of the green brainline (here between nodes #3 and 4 for visibility) with a supraliminal segment. As a result synchronization is lost at the end of the red and green brainlines.*

It has been well established that most schizophrenic patients hear voices and fMRI studies have revealed that their auditory cortex is, indeed, activated, as also evidenced during auditory hallucinations of otherwise normal subjects [54]. We all perceived internal voices (auditory imagery), when we think, read, etc., but we know they are ours. To illustrate the concept let us consider a schematic (obviously over simplistic) view of the normal situation where internal thoughts are produced unconscientiously in an area of the prefrontal cortex (event 0 in node 0), flows to the frontal cortex (event 1, node 1), then the auditory cortex (where "voices" are recreated) (event 2, node 2), continuing to language areas (which identify their lexical "content") (event 3, node 3) and back to the frontal cortex (event 4, node 4) which associates the inner voices with the initial thoughts in full consciousness. Taking now the view within the referential of the flowing activation (axes (c*t' and x') the pathological occurrence of a segment with supraliminal speed between nodes 1 and 2 will result in event 2 arising before event 1. In other word the recreation of voices (in the auditory cortex) occurs *before* initial thoughts reach node 1 (frontal cortex) (Fig.5b).

The association between internal thoughts and imagery auditory activity is then broken, resulting in the inner voices being not recognized as internal but external. In short, schizophrenia might be considered as a disconnection disorder with *oneself*.

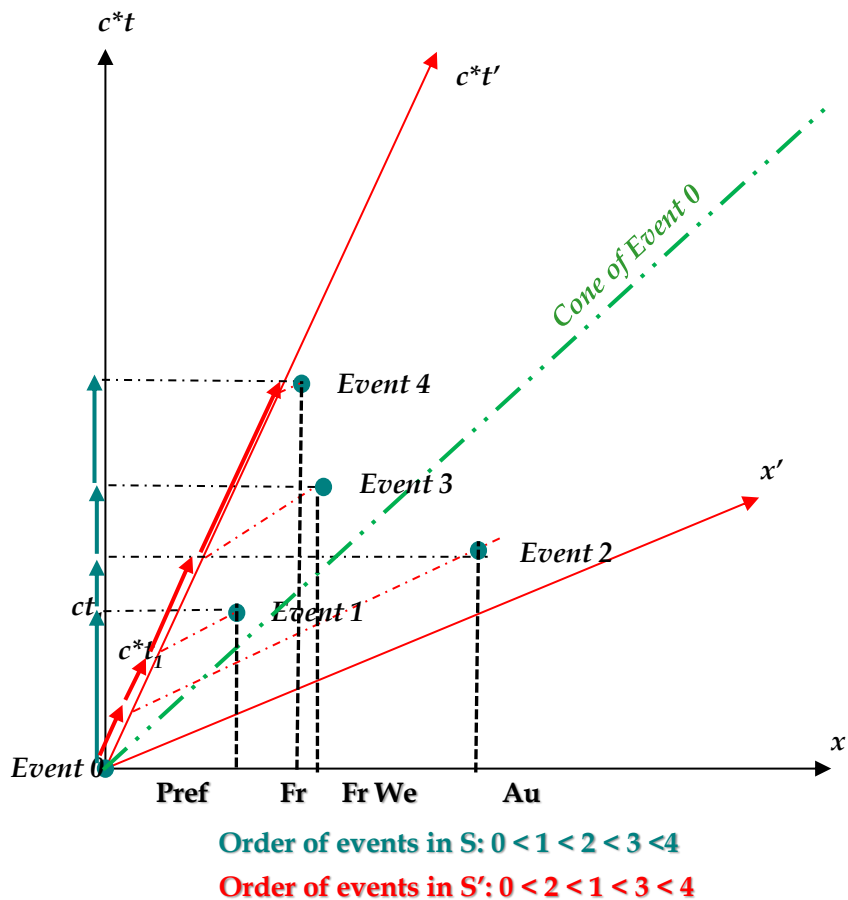

**Figure 5b: Reversal of event time order in schizophrenia**
*Reducing the spacetime in 2 dimensions one may look at the proper timeframe experienced from event 0 (1 for space, c*t, and 1 for time, x) and by a propagating action potential (c*t', x'). Here the segment between nodes #2 and 3 was made supraliminal. As a result, the order of event for the action potential is 0, 2, 1, 3 and 4 while it remains 0, 1, 2, 3, 4 for an observer linked to event 0.*

A similar situation may occur with "déjà vu" situations which most of us have experienced. Our perception of stimuli from environment (event 1) is dynamically compared to memorized patterns (2) for recognition and then stored (event 3). If for some reason storing occurs faster than event 2 the stimuli conditions will appear as already existing in memory, hence the "déjà vu" impression.

## 5. Relativistic pseudo-diffusion brain model
### 5.1. Brain activity pseudo-diffusion concept

As the next step we now consider (pseudo) diffusion as a vehicle to build a global model of the working brain. In short, neural activity would be considered as "bouncing" between brain nodes while following pathways determined by the 4D brain spacetime geodesics.

The diffusion and random walk concepts, as set by Einstein [6] have given rise to may fecund models across disciplines beside molecular physics and chemistry, up to finances [55] and recently cosmology [56]. For instance, diffusion MRI (and DTI) has been developed to obtain brain functional and architectural information based on its tissue microstructure. As an example of "pseudo-diffusion" let us mention blood microcirculation (flow) in capillary vessels in the brain or other organs. Blood microcirculation has been very successfully modeled as a random walk associated with a pseudo-diffusion coefficient within the IntraVoxel Incoherent Motion (IVIM) MRI concept [57-58]. The pseudo-diffusion coefficient of flowing blood is about one order of magnitude larger than the molecular diffusion coefficient of water in brain tissue (which is about $1\ 10^{-9}$m²/s). Although flow of blood in capillaries is a completely deterministic process, this pseudo-diffusion model assumes, by principle, a stochastic distribution of blood microcirculation. This apparent contradiction is resolved when considering that blood flow pseudo-randomness *in time* results, indeed, from a pseudo-random organization of capillary segments *in space* within brain tissue, illustrating how the diffusion concept can be used as a device to mix up space and time. We might not exclude that this sight has been in Einstein's mind given the publication of his famous papers on diffusion and special relativity within a few months of each other (though see below).

When observing 3D random walks (real or simulated) of particles one is often surprised by the spatio-temporal patterns described by the particles which often remain within a short domain before jumping to a more remote location where they will remain again for a short while before extending again to another remote location (Fig.6). Indeed, those patterns may well mimic how neural activity progresses locally within adjacent nodes or within neural circuits with short-range connections before extending to remote nodes through long-range connections. As with the blood microcirculation pseudo-diffusion model this view does not imply at all that the network is "random" (as considered for instance in the Erdos-Renyi random network model [59]), though some randomness is very likely present in the network, especially in early times of life. Indeed, randomness is part of small-world models [60] which accommodated a certain degree of disorder in an otherwise deterministic organization within the network. Transition between geometric (deterministic) networks and random networks can also be created using random geometric graphs depending on critical model parameters (eg giving connections a "cost" depending on their physical distance) [61] or emerge from neural correlations using the "clique topology" approach [62]. More generally, the transition between a deterministic and a random vision emerges when we lose tracks of the minute details of the many deterministic events, in other word when we blur our vision. In a deterministic world the future and the past are completely interrelated and cannot be, in principle, disentangled (as laws of physics are reversible in time). Time issues (and irreversibility) appear with the blurring (though the opposite might be true when considering Quantum Mechanics).

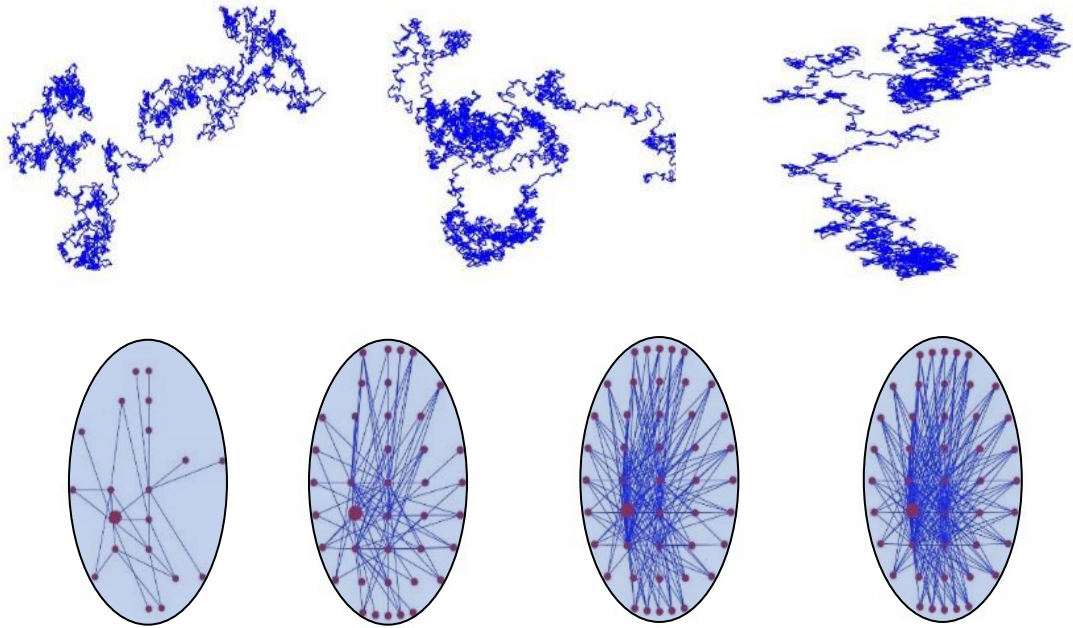

**Figure 6: Brain pseudo-diffusion model**
*Top row: Simulated unrestricted random walks with increasing diffusion times.*
*Bottom row: Simulated random walks within a node network with boundaries in space (oval shape representing a slice through the brain) and time (speed limit). Random walks originated from the largest red circle.*

If D* is the pseudo-diffusion coefficient associated with a pseudo-random flow of neural activity within the brain the "diffusion" distance, r, covered within the nodes network after a time t should be given, in principle, by Einstein's diffusion equation [6]:
$$<r^2> = 6 D^* t \qquad [3]$$

A first correction to Eq.[3] is necessary when considering that the spatial dimensions of the brain are limited. This is not a difficult task as there is an extensive literature on *restricted diffusion* imposed by geometric barriers [9].

### 5.2 Relativistic diffusion

However, a problem which immediately arises when considering the diffusion equation Eq.[3] is to realize that it was derived by Einstein assuming that velocities of "particles" undergoing a diffusion process or a random walk can be infinite [6]. This means, very surprisingly, that Einstein's 1905 articles on diffusion and special relativity (which makes the light speed as a Universe physical limit as a principle, [49] are, in fact, in conflict. This "oversight" becomes an issue for our pseudo-diffusion model as, in the brain, there is, indeed, a physical speed limit, c*, for neural activity flowing within the brain connectome.

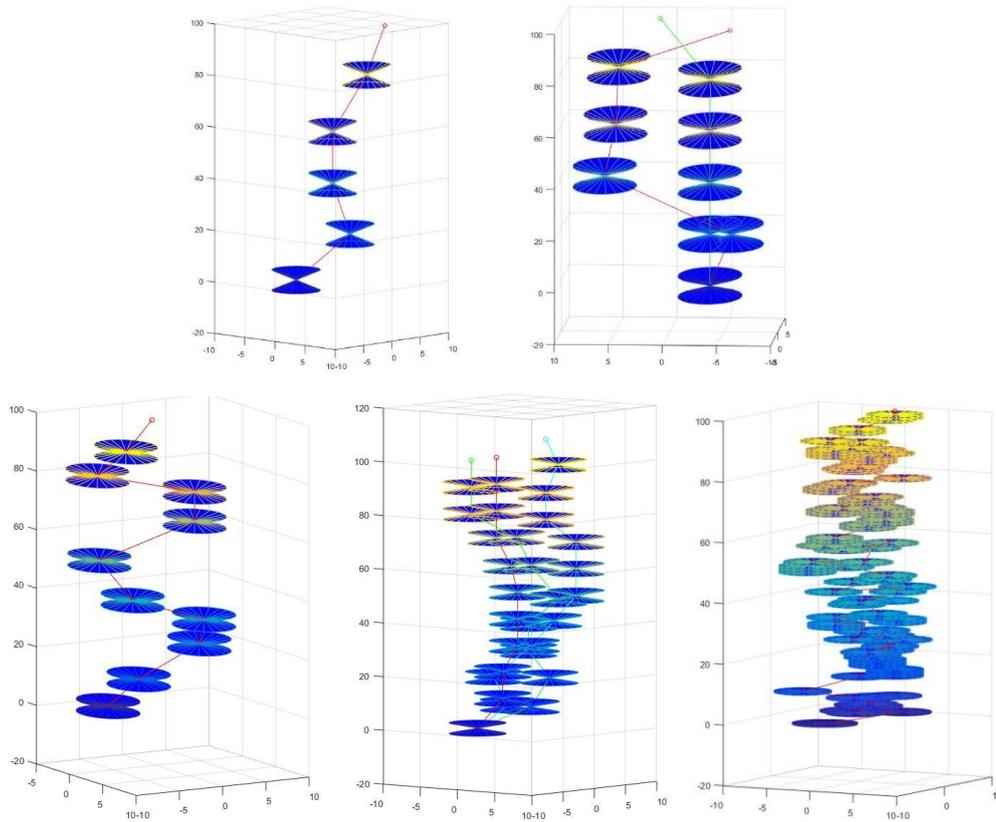

**Figure 7: Minkowski diagrams of relativistic pseudo-diffusion in brain spacetime**
*Examples of brainlines issued from a given event generated using the relativistic pseudo-diffusion model using different speed limits c\*.*

The existence of a speed limit (whether in the Universe or the brain) requires Eq.[3] to be corrected to take into account "relativistic" diffusion effects. Relativistic diffusion has been considered in the past using the telegraph equation [63], but this approach suffers limitations and there was no commonly accepted solution until recently [64-66], as diffusion has become a topic of interest in high energy physics and astrophysics. Dunkel et al. have derived a generalized Wiener process for particle traveling between *spacetime* events, which can elegantly accommodate both relativistic and nonrelativistic processes, and allow to derive a relativistic diffusion propagator within Minkovski spacetime framework:

$$p_w(\bar{x}|\bar{x}_0) = \mathcal{N}_w \int_{a_-(\bar{x}|\bar{x}_0)}^{a_+(\bar{x}|\bar{x}_0)} da\ w(a) \qquad w(a) \equiv \exp[-a/(2\hat{D})] \quad [4]$$

where $p_w$ is the transition PDF of the Wiener process, $N_w$ a normalization constant, $a_-$ and $a_+$ the minimum and maximum action values along the path (such as $a_-$ is the negative Minkowski distance of the spacetime events $x_o$ and $x$) and $w(a)$ a weighting function (eg $w(a) \equiv \exp(-a/2D^*)$ ).

Obviously, the resulting equations have been developed for the physical world using the famous "Lorentz factor" and an extension to our brain pseudo-diffusion model must remain at the analogy level at this stage (the advantage of Dunkel's model is that it can accommodate a very large class of diffusion functions *w*). Nonetheless, the overall result, that is the relationship between the diffusion distance covered by "particles" traveling with a speed limit and time, might be relevant, at least qualitatively (Fig.7, 8). Extrapolating for the brain one obtains that the pseudo-diffusion distance among brain nodes becomes shorter than expected from Eq.[3]:

$$<r^2> = 6 \gamma D^* t \quad\quad\quad [5]$$

where $\gamma$ ($<=1$) now depends on ($c^{*2}t/D^*$). In other words, the pseudo-diffusion distance appears reduced due to relativistic effects when ($c^{*2}t$) is small or $D^*$ is large (Fig.7,8). The metric tensor used for Eq.[2] should now be rewritten as:

$$g_{\mu\nu} = \text{diag} ((6D^*)^{1/2}/t, -1, -1, -1) = g_{\nu\mu} \quad\quad\quad [6]$$

### 5.3 Implication for brain states

As it appears clearly from simulations (Fig.7, 8) the potential number of nodes which can be connected (connectivity level) directly depends on the connectome speed limit, which, as we have seen, increases with brain maturation. One may consider that this setup gives the richest, most effective configuration (wide cones) allowing strong global coupling (with high integration and low segregation) when activation flows near full capacity, as under exposure to hallucinogens [73]. On the other hand, if the connector speed limit is decreased, for instance for pathological reasons, connectivity will be reduced (narrower cones), resulting in weak global coupling (low integration and high segregation). This reduction might be "static", resulting from faulty anatomical connections, or "dynamic" in relation with special brain functional conditions, for instance during anesthesia or in vegetative states. The normal (conscious) state lies in between, with an optimal level of coupling balance between segregation and integration.

Hence, this framework may also provide some complementary perspective on recent fMRI findings linking consciousness in humans and non-human primates with the richness of brain signal dynamics reflecting a large repertoire of functional configurations carried out by strong coordination and long-range temporal correlation between distant cortical regions [68]. To the contrary the brains of minimally conscious or unresponsive patients showed primarily a less efficient pattern of more temporally rigid, low interareal phase coherence almost limited to the largest structural pathways. Patients with in-between stages of consciousness exhibited intermediate connectivity patterns, as also found with this relativistic pseudo-diffusion model (Fig.8). Similarly, during anaesthesia performed cortical long-range connections are temporally disrupted both in time and space [69-70]. Overall, unless there are anatomical faulty connections (see below) the present model suggests that there is a continuum of states and that the brain nodes status (wide or narrow cones) varies or fluctuates between different connectivity levels, depending on attention the subject gives to environment.

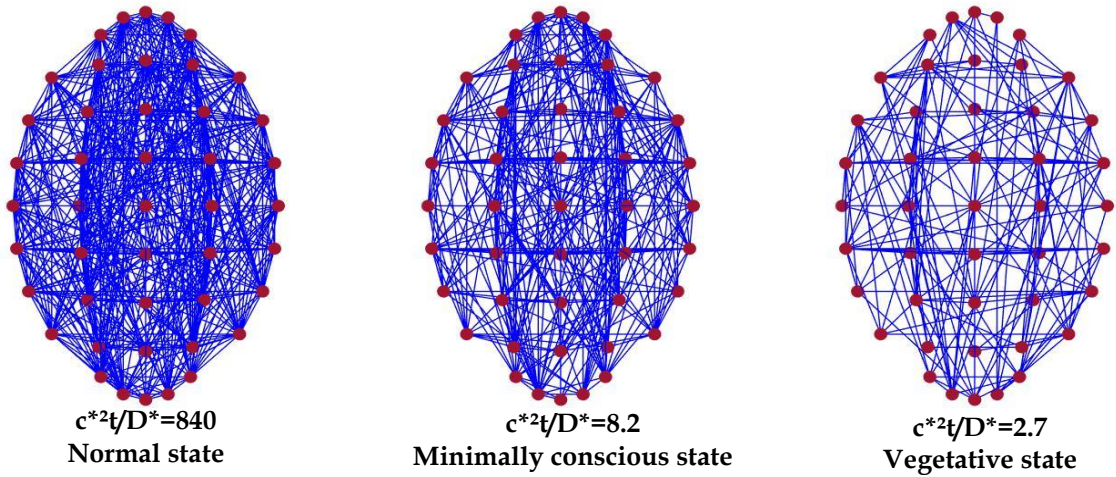

**Figure 8: Relativistic pseudo-diffusion model for different patterns of brain activity**

*Simulation of brain connectivity based on the relativistic (restricted) pseudo-diffusion model. All red points were considered as starting points with equal probabilities (equivalent to a "resting-state" status). Giving a 100ms time window the number of nodes getting connected highly depends on the network speed limit, c\*, and the resulting pseudo-diffusion coefficient, D\*. Interestingly the figures highly resemble connection patterns observed in the brain of patients in normal, minimally conscious and vegetative states (see for instance Fig.1 in [68]).*

## 6. Curvature of brain spacetime

### 6.1. Link between the pseudo-diffusion coefficient and the brain network functional structure

It remains to see what $D^*$ may look like in this model. With the perfusion pseudo-diffusion model (see above, Le Bihan 1988) the associated pseudo-diffusion coefficient was simply "imported" from Einstein's random walk equation, $D^* = \langle l \rangle \langle v \rangle / 6$, where $\langle v \rangle$ the average speed of diffusing particles and $\langle l \rangle$ the average distance between two successive collisions. For perfusion $\langle l \rangle$ was taken as the average capillary length (about 50μm in the brain) and $\langle v \rangle$ as blood velocity in the capillary network (around 1-1.5mm/s), resulting in $D^* \sim 10 \cdot 10^{-9} m^2/s$ [57]. This theoretical value has been found later remarkably close to experimental values (8-20 $10^{-9} m^2/s$) obtained in the normal brain and various pathological conditions by IVIM/diffusion MRI [58], validating the approach despite its crudeness and that the values for $\langle l \rangle$ and $\langle v \rangle$ are order of magnitudes out of the ranges relevant to molecular diffusion and initial Einstein's theory. For the brain activity pseudo-diffusion model, as $\langle l \rangle$ and $\langle v \rangle$ are intimately linked in the spacetime one may prefer another approach to define $D^*$. By considering pseudo-diffusion in a network of N nodes, the density of which is ρ and the "cross-section" σ (a parameter representing the strength of a node to connect to other nodes) one may define a "mean free pseudo-diffusion path", $\lambda$, by analogy to the kinetic theory, as $\lambda=1/(\sqrt{2}\rho\sigma)$. The pseudo-diffusion coefficient, $D^*$, can then be defined as

$$D^*=\omega\lambda^2/2= \omega/4\rho\sigma \qquad [7]$$

where $\omega$ would be the average "collision" (firing) rate within the network. In other word $D^*$ could be seen as a marker of the level of the segregation/integration balance, a balance between local information processing within clusters (low $D^*$, resulting from high $\rho\sigma$) and global information transmission across long-distance connections (high $D^*$, resulting from high $\omega$). Locally (within a node) $D^*$ would give an indication of its strength (level of activity, $\omega$) and the number of connections it could establish ($\sigma$). To some extent this view is comparable to Rentian scaling which describes a relationship between the number of nodes in a volume and the number of connections crossing the boundary of the volume [71].

**6.2.** Hyperspace of possible network configurations associated to different geodesics

Until this point, we have considered events (and their associated cones) independently of each other. More generally, we have seen sets of events in the brain patched as a 4-dimensional pseudo-Riemannian space which represents the brain spacetime, such that squared displacements, $ds^2$, between neighboring events can be expressed as $ds^2= g_{\mu\nu} dx^\mu dx^\nu$. Neural activity (in the form of action potentials) is thus expected to follow locally "straight" geodesics ("brainlines") in the brain spacetime which we have considered as "flat" until this stage.

With General Relativity Einstein showed that the metric of the Universe spacetime was related to the stress-energy tensor of the sources it contents (including gravity) through field equations, giving the spacetime a curvature [2]. By analogy, we might consider that the "trajectories" (brainlines) followed by neural activity in the brain 4D-spacetime are determined by its metric, which, in turn, is determined and shaped (curved) by its sources, that is by brain nodes activity ("energy"), and existing connections ("field") between them (Fig.9). Note that the situation is a little bit easier for the brain than the Universe where the mass concept had first to be converted to energy (and then a stress-energy tensor) through the Relativity theory [72]. However, reciprocally one may also virtually associate node activity (energy) level to a "neural mass" (actually, neuronal activity might be really accompanied by changes in local mass considering that neural swelling and influx of water occur, notably within dendritic spines, upon activation [73], but this is out of the scope of this essay).

Following Einstein's approach to establish his field equations [2] we propose to derive the curvature of the brain spacetime, R, through its metric, $g_{\mu\nu}$, from its energy content in tensor form (reduced to 10 independent components considering the tensor is symmetric), as:

$$R_{\mu\nu} -1/2\ R\ g_{\mu\nu} =-kT_{\mu\nu} \qquad [8]$$

where here $R_{\mu\nu}$ represents the symmetric Ricci tensor, R the Ricci curvature scalar (0 in the absence of brain activity), $g_{\mu\nu}$ the brain spacetime metric, k a constant ($s^2kg^{-1}m^{-1}$) to be determined and $T_{\mu\nu}$ the stress-energy rank 2 tensor describing the distribution and flow of activity in a region of brain spacetime. $T_{00}$ is the local energy density in the nodes while the other terms are the rate of energy flow in spacetime around the nodes. Considering the pseudo-diffusion model, to some approximation $T_{\mu\nu}$ would be linked to $D^*$, with at any brain spacetime event $T_{00}\sim\rho\sigma$ (local node energy density) and $T_{ii}\sim\omega$ ("pressure" or the energy being transferred per unit area and unit of time in all directions by the "thermal motion" representing the pseudo-diffusing activity flow).

Solving Eq.[8] means finding the metric tensor $g_{\mu\nu}$ for a given energy configuration $T_{\mu\nu}$ to obtain the brain spacetime curvature and then derive the related geodesics (brainlines) through Eq.[2]. Unfortunately, as with General Relativity, solving this set of 10 simultaneous, non-linear, second-order partial derivative equations condensed in Eq.[8] is not straightforward at all and might even be impossible unless some assumptions are made (as simplified metrics which have been proposed as approximate solutions for General Relativity), with in addition the complexity of the pseudo-diffusion model, which has never been considered. Fortunately, nowadays one may use computer simulations (Fig.9).

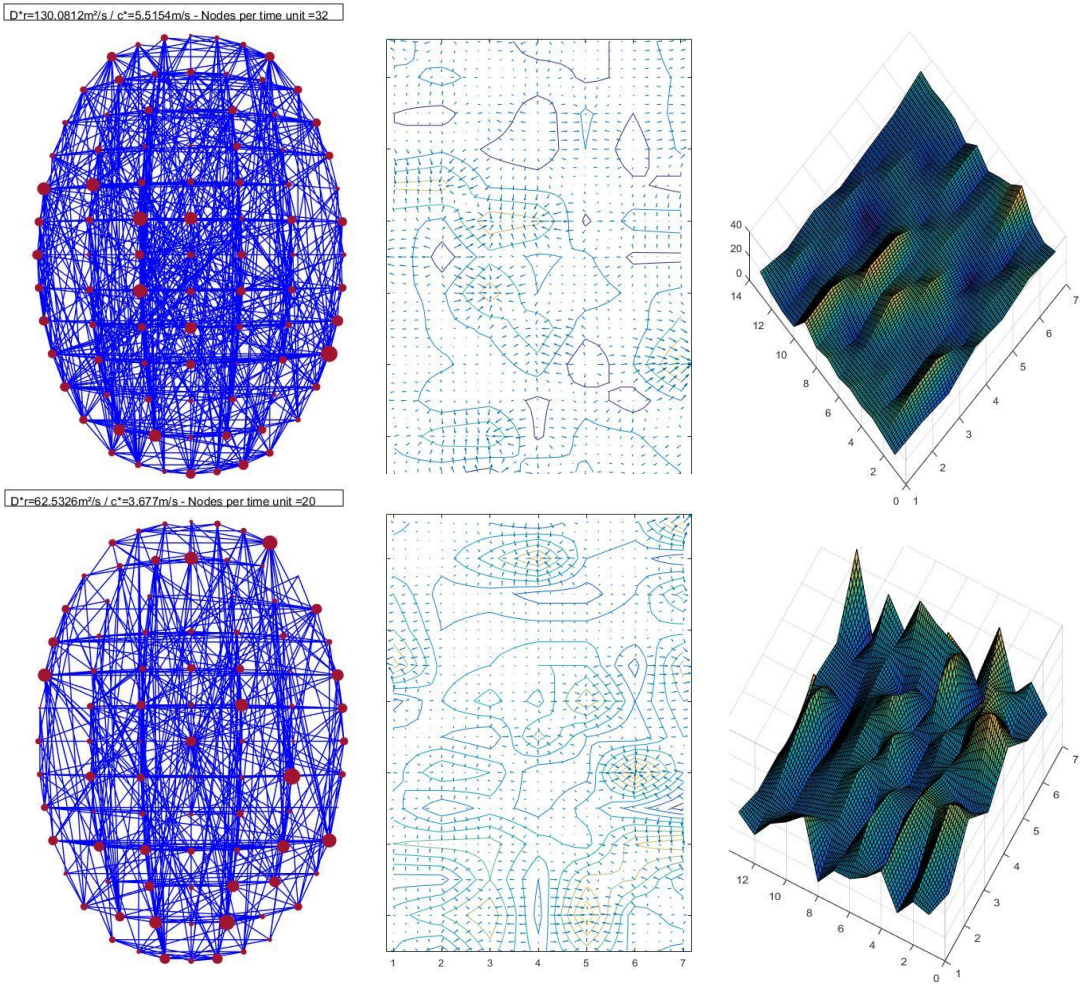

**Figure 9: Curved brain spacetime and related geodesics**

*Simulation of connections in a brain node network (left) using the relativistic restricted pseudo-diffusion model (left). Some nodes were associated with a higher probability to get hit (the size of the red dots corresponds to the number of times a node was hit during 100ms). The connection pattern may be represented using vectors pointing to the nodes according to their*

*degree of "attractivity" (energy/field map, center) or equivalently using a 3D display of the associated curvature of the brain spacetime (activation flowing toward "valleys")(right).*

Generally, within this framework neural activity now becomes a part of brain spacetime geometry which plays the role of a field. At this level, no assumption is needed for what is actually happening within each node: Neural activity does not "feel" nodes activity directly but through the connections the nodes have built in their neighborhood. As a result of this curvature the event little cones we have elaborated above will no longer be "parallel" to each other everywhere, but a brainline will nonetheless be within the cones at each of its points. Cones can become quite tangled and two of them issued from a single event may "meet" again at a later stage (something known as "gravitational lensing" in cosmology). It is even conceivable that successive cones could be tilted in such a way that they make loops: A brainline trajectory toward the future ends up at the event where is started, that is in the past, blurring "cause" and "effect" (or "memory" and "intention" to take a more anthropomorphic comparison). While this possibility might appear puzzling for the Universe to our mind formatted through compulsory education to Newton's absolute time principle it might well exist in the brain and support how we can revive past experiences and memory.

The importance of the brain's inner time and space to support neural activity to consciousness has been reported earlier [74], but as separate entities. The proposed view allows gray matter (node) "activity" and white matter (edge) "transmission" to be merged into a unique framework to describe the natural flow of neural activity within the brain within a spacetime referential. In sum, the "geometry" of the brain 4D-spacetime is determined by the activity of its nodes. In other word nodes (especially hubs and basal ganglia, such as the thalamus, van den Heuvel 2013) act in a similar way as "gravitational masses" in the physical spacetime to explain "coupling". Neural node "masses" (especially hubs which have been considered as neural "sources and sinks") curve brain spacetime, activation flow increases node masses. Hence, our thoughts can be seen as carried by continuously changing curved brain spacetime landscapes (Fig.9). By combining fMRI with pulsed optogenetic stimulation paradigms to probe the spatiotemporal dynamics and functions of the well-defined topographically-organized somatosensory thalamo-cortical network Leong AT et al. [75] revealed that unique long-range propagation pathways are, indeed, dictated by distinct neural activity temporal patterns initiated from the thalamus.

### 6.3 Implication for normal and diseased brain function

The architecture of the brain spacetime should not only impact our understanding of cognition but also our understanding of neurological or psychiatric disorders (or at least their clinical expression), now seen as disorders of the brain spacetime, of combined structure and function, as we will see. Based on the above description an event occurring at a brain node will be represented by a cone. The node receives inputs from a limited number of nodes (reverse causality) based on the connectome speed limit. Conversely, the node will send inputs to a limited number of nodes (forward causality).

### 6.3.1 The role of attention and priming

We give here some examples of previous fMRI studies which might be reinterpreted under the light of the proposed model.

In a masking paradigm in which subliminal versus supraliminal letter strings were presented to subjects while they focused attention on another subsequent, highly visible target word [76]. Subliminal strings elicited responses confined to bilateral posterior occipito-temporal cortices, without extension into frontal and parietal cortices. However, supraliminal stimuli evoked more extensive responses in a widely distributed set of parieto-frontal areas. Furthermore, only supraliminal primes caused phonological repetition enhancement in left inferior frontal and anterior insular cortex. In light of the present model we might interpret the activation of multiple distant areas beside occipito-temporal responses only when the brain spacetime is curved enough by attention, as present only when normal (unmasked) stimuli are presented. Otherwise, the parieto-frontal processing network is not activated, and stimuli remain unconscious.

More generally, priming effects might be explained in light of the suggested framework. Exposure to stimuli or environmental events, even unconsciously, are known to profoundly affect the speed, intensity or even he kind of responses we make to other stimuli or environmental events, up to our reasoning and emotions (see for instance, [77]). Those priming events may be seen as curving our brain spacetime, so that subsequent events or activation patterns simply follow the geodesics set by the priming events.

In a very interesting fMRI study Hesselmann et al. [78] investigated brain activity of subjects who reported perceptual decisions on the famous ambiguous Rubin's vase-faces picture which was briefly presented during 150ms followed by a 250ms noise mask. It was found that subjects perceived faces instead of the vase when an activity in the fusiform face area (FFA), a cortical region preferentially responding to faces, was present *prior* (by a few seconds) to the visual stimulation. The absence of prior activation in this area resulted in the perception of the vase. Because of the variation between trials the results were interpreted as an effect of spontaneous fluctuations and not from those of cued attention or induced mind sets. While the "spontaneous" nature of ongoing activity might be interpreted in the framework of some computational models from theoretical physics [79], the FFA activity might also correspond to concurrent, unconscious processes occurring at the time in this node. In any case, this experiment shows that neural activity in a specific node impacts on how we make up our mind during subsequent perceptual decisions on sensory inputs. One interpretation has been proposed within the framework of the prediction error and free energy suppression theory [80]. However, we might as well consider that, among the two brainlines (from the visual input of the image to the expression of a decision) corresponding, respectively, to the perception of faces or a vase, the one which corresponds to the shortest (straightest) geodesic will prevail when the stimulus enters brain spacetime. The local spacetime configuration (curvature), and associated geodesics (pathways requiring the minimal energy cost), will be determined by the amount of energy (activity) in local nodes, that is the FFA in this context: Activity in this node will lead the brainline associated with the visual stimulus to pass through the node, resulting in the perception of faces. Subsequently, if more time is given to perceive the stimulus, one may decide at will whether to see a vase or faces, which would imply a reconfiguration of the brain spacetime and its local curvature through cued attention.

Similarly, Sadaghiani et al. [81] showed using fMRI that the connectivity state within large-scale networks before playback of a faint sound predicted whether the participant was going to perceive the sound on that trial. Connectivity states preceding missed sounds showed weakened modular structure, in which connectivity was more random and less organized across brain regions. Connectivity status and brain spacetime curvature can be controlled overtime by attention and learning. Mukai et al. [82] showed that repeated experience with a visual stimulus can result in improved perception of the stimulus through learning (perceptual learning). Compared to non-learners, learners showed first an increase in activity in both visual cortex and the fusiform gyrus, as well as cortical regions associated with the attentional network (intraparietal sulcus, IPS, frontal eye field, FEF, and supplementary eye field), then progressively reduced activation with some correlation with the magnitude of performance improvement progressively reduced activation. Interestingly, over the course of training, the functional connectivity between IPS and FEF in the right hemisphere with early visual areas was stronger for learners than non-learners, suggesting that attention may facilitate this process through an interaction of attention-related and visual cortical regions.

In addition, visual perceptual learning has been shown to decrease latencies in some visual evoked potential components, notably N1, negatively correlated with improvements in behavioral performance [83]. To the contrary, attentional effort increases N1 latency [84]. Those results support the idea that attention and learning have an impact on the curvature of the brain spacetime between early and late visual processing areas, modifying the "bending" of the brainlines between those areas, and, hence, the observed time courses between events.

### 6.3.2 Consciousness and brain stimulation

Another field of application for this model is consciousness. In a conscious state brain activity occurs optimally as each node can be functionally connected to many others, in the past and the future. Considering the whole brain we may also see this state corresponding to a high degree of curvature in brain spacetime which allows activity to flow between nodes. If, for some reasons, activity of critical nodes declines (eg, thalamus) the brain spacetime will tend to become less curved, more flat, and nodes will become somewhat isolated (reduced connectivity), although local function could still exist. This might occur in patients in vegetative states: Patients emerging from coma appear to be awake but show no sign of awareness. In some way, the situation is comparable to a neighborhood where each house would have internal wireless connections between rooms, but no connection to Internet. In a famous study Owen et al. [85] carried out using fMRI in a young lady in a vegetative state the authors showed that the patient exhibited activation patterns in several specific brain regions identical to those observed in normal subjects when "asked" to imagine playing tennis or navigating inside her house, although there was no speech or motor response.

According to the present model one may, thus, expect some day to be able to get a patient out of vegetative state by increasing, through some means, the curvature of this patient brain spacetime. To do so we would have to act on specific brain nodes, especially the thalamus which is known to be involved in awake/sleep states status, providing locally energy (or "mass"). As a proof-of-concept study we have successfully been able to awake rats under anesthesia by manipulating the thalamus central median (CM) nucleus [73]. With diffusion MRI we showed that this nucleus was functionally connected to a few important other brain locations, such as the cingulate and somatosensory cortex. Electrical stimulation of the CM nucleus led the animals to transiently wake up. Infusion in CM of furosemide, a specific neuronal swelling blocker, led water diffusion to increase further locally and increase the electrical current threshold in CM necessary for the awaking of the animals. Oppositely, induction of cell swelling in CM through infusion of a hypotonic neutral solution led to a local water diffusion decrease and a lower current threshold to wake up the animals. Together, those results strongly suggest that the modulation of neuronal swelling (linked to the degree of local activity or energy, and D*) in CM plays a significant role in allowing or not the animal to wake up, although under the condition of anesthesia. In view of the proposed model CM would act as a "mass" curving the animal brain spacetime around the thalamus and bending geodesics to increase the amount of communication between brain regions otherwise isolated as a consequence of anesthesia.

Deep Brain Stimulation which is increasingly being used in clinical patients to alleviate symptoms of some neurological disorders (notably Parkinson's disease) and potentially mental illnesses might also benefit from considering the new framework of the brain spacetime. More generally, this framework might be useful when considering the internal or external control (for instance using electrical stimulation or transcranial magnetic stimulation) of brain networks not only for medical treatments, but also potentially to enhance performance or cortical processing abilities [86-88].

### 6.3.3 Social interactions: Application to autism and Alzheimer's disease

Koike et al. [89] have shown how the limbic mirror system, including the anterior cingulate cortex (ACC) and anterior insular cortex (AIC), both of which are critical for self-awareness, and the cerebellum (left cerebellar hemisphere and vermis) are essential when individuals share affective and mental states through eye contact, a critical element of social interaction. To do so they conducted a hyperscanning fMRI study involving on-line (LIVE) and delayed off-line (REPLAY) conditions on pairs of individuals, analyzing pairwise time-series data for eye-blinks, considered as representing individual's attentional window toward the partner. Relative to the REPLAY condition, the LIVE condition was associated with greater activation of the above structures. In light of the present framework one may see those structures when activated as curving the brain spacetime of the individuals so to establish the necessary connections to support social interaction.

This study has important implications to understand social contingency, sense of self and interaction with others, and ultimately joint-attention deficits in autism. There are many reports showing that autism might be considered as a disconnection disorder with *others* (as opposed to *oneself* for schizophrenia). Indeed, DTI studies have outlined faulty connections (mainly in fSCS, IFOF, uncinate and arcuate fasciculus which are known to be involved is social interactions) with abnormal growth is Autistic Syndrom Disorders (ASD) infants, to the extent that DTI data might predict ASD in toddlers [90]. In those individuals, because of the low speed limit present in those faulty pathways, timely connections might not be established properly, resulting in the ASD symptoms.

Let us see now how the brain spacetime framework could be used to account for such social interactions. To do so we first consider the constant, $\Lambda$, (known as the cosmological constant) which Einstein added to his field equations to account for the static nature of the Universe [91]. He then considered this addition as the "greatest blunder of his life" when it was found that the Universe was not static but expanding. With this constant Eq.[8] becomes:

$$R_{\mu\nu} - 1/2\ R\ g_{\mu\nu} + \Lambda\ g_{\mu\nu} = -kT_{\mu\nu} \qquad [9]$$

This equation can be rewritten as:

$$R_{\mu\nu} - 1/2\ R\ g_{\mu\nu} = -k\ (T_{\mu\nu} + T'_{\mu\nu}) \qquad [10]$$

with $T'_{\mu\nu} = (\Lambda/k)\ g_{\mu\nu}$.

As a spectacular upturn $T'_{\mu\nu}$ (and $\Lambda$) have been reconsidered later to account for the presence of "dark energy" (density and related pressure) in the Universe. However, there is another possible interpretation for $T'_{\mu\nu}$ which could be relevant for the brain spacetime. This additional term could be seen from a totally different perspective, as an external source of energy (and momentum) which might impact (curve) one's brain spacetime and its metric, namely the interaction with brain spacetimes of others. While one generally admits that there is only one Universe we should appreciate that there are 7 billion brains on Earth, all of them interacting locally through many different forms of languages, and now also at large scales through libraries, internet and social media. Hence, there would be not a single, but multiple occurrences of the stress-energy tensor $T'_{\mu\nu}$ coexisting and varying in space and time to impact a given brain spacetime, and reciprocally (Fig.10). The brain spacetime field equations and the brain spacetime metric for an individual interacting with *N* others would then take the form of:

$$R_{\mu\nu} - 1/2\ R\ g_{\mu\nu} = -k\ \Sigma_{i=0,N}\ T^i_{\mu\nu} \qquad [11]$$

The pseudo-diffusion model can easily be rescaled across *N+1* brain spacetimes. Indeed, the possibility to transmit activity or information across brains spacetimes, which may lie miles and/or centuries away, as supported by Eq.[11] (a sort of "social interaction tensor" equation) may be what makes humankind unique among species living on Earth.

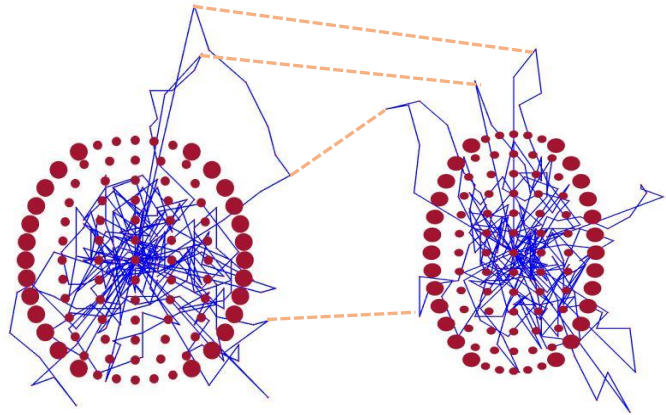

**Figure 10: Curved brain spacetime and related geodesics**
*Schematic view of 2 brains figuring how brain activity issued by nodes from one brain curved spacetime could interact with nodes from another brain, curving its spacetime, and reciprocally, through body interactions in space or time (dotted orange straight lines).*

## 7. Conclusion

Even the best of physical theories are only an approximation of the truth. Such theories, systemizing experimental facts, are just human imaginative entities which assert not so much what nature is, but rather what it is like to help us make predictions, most often using mathematical expressions (often through differential equations). In the case of the brain modeling lies at the borders of biology, physics and even philosophy. With the arrival of outstanding instruments, such as ultra-high field MRI scanners, brain theories can be falsified more easily, making them more science than speculation. The global framework we propose aims at specifying how time may be intermingled with the spatial organization of the brain resulting in a 4 dimensional spacetime sculpted by brain nodes activity. It is not a unified theory of everything but a coherent physical and biological system at global level to merge gray matter "activity" and white matter "transmission" into a unique framework to describe how neural activity flows within the brain. This spacetime is curved by the activity or energy ("mass") present in brain nodes made of gray matter clusters. The activity generated by those nodes flows along connections (connectome) following "brainlines" which are geodesics in this brain curved spacetime. In other word geodesic tracks representing activity flow within the brain follow specific connections which dynamically depend on the changing gradients resulting from the activity potential. To paraphrase JA Wheeler one may conclude that brain spacetime tells activity how to flow while activity tells brain spacetime how to curve.